\documentclass[a4paper]{kluwer}

\begin{document}
\begin{article}
  
  \begin{opening}
    \title{Merging Massive Star Clusters as Building Blocks of
      Dwarf Galaxies ?}
    \runningtitle{Merging Massive Star Clusters}
    \author{M. \surname{Fellhauer}}
    \author{P. \surname{Kroupa}}
    \institute{Inst.\ Theor. Phys. \& Astrophys., Univ.\ Kiel,
      Germany} 
    \date{\today}
    \begin{abstract}
      Recent spectroscopic observations of galaxies in the
      Fornax-Cluster reveal nearly unresolved `star-like' objects
      with red-shifts appropriate to the Fornax-Cluster.  These
      objects have intrinsic sizes of $\approx 100$~pc and
      absolute B-band magnitudes in the range $-14 < {\rm M}_{\rm
        B} < -11.5$~mag and lower limits for the central surface
      brightness $\mu_{\rm B} \ge 23$~mag/arcsec$^{2}$ (Phillipps
      et al.\ 2001, Hilker et al.\ 1999), and so appear to
      constitute a new population of ultra-compact dwarf galaxies
      (UCDs). Such compact dwarfs were predicted to form from the
      amalgamation of stellar super-clusters ($=$ clusters of
      star clusters; not to confuse with super stellar clusters
      (SSC)) by P. Kroupa (1998), which are rich aggregates of
      young massive star clusters (YMCs) that can form in
      collisions between gas-rich galaxies.  Here we present the
      evolution of super-clusters in a tidal field.  The YMCs
      merge on a few super-cluster crossing times.
      Super-clusters that are initially as concentrated and
      massive as Knot~S in the  interacting Antennae galaxies
      (Whitmore et al. 1999) evolve to merger objects that are
      long-lived and show properties comparable to the newly
      discovered UCDs.
    \end{abstract}
  \end{opening}

\section{Introduction}
\label{sec:intro}

Spectroscopic surveys of the Fornax-Cluster revealed 5 compact
objects which have spectra typical for late-type metal-rich and
old stellar populations, are marginally resolved and have
red-shifts comparable to the Fornax-Cluster, ruling out either
faint background galaxies or foreground stars.  Four of
these may be globular clusters at the very bright end of the
luminosity function (Mieske et al. 2001) but one is
resolved with a diameter of $\approx 300$~pc.  Phillipps et
al. stated that these objects form a new class of objects, namely
Ultra-Compact Dwarf Galaxies (UCD).  The UCDs have intrinsic
sizes of around 100~pc, absolute B-band magnitudes of $-14 < {\rm
  M}_{\rm B} < -11.5$~mag and lower limits (due to failure to
resolve their cores) of the central surface-brightness $\mu_{\rm
  B} \geq 23$~mag/arcsec$^{2}$.  Further analysis of
photographic plates show no sign of low-luminosity envelopes
around these objects which rules out the possibility that they
are nucleated dwarf ellipticals (dE,N) with faint envelopes.
Therefore, these objects are either extremely compact dwarf
galaxies or extremely large and massive ($10^7-10^8\,M_\odot$,
assuming $M/L_{\rm B}=3$) star clusters.  In the $\mu_{\rm
  B}-M_{\rm B}$ diagram they fall in the empty region between
`ordinary' dwarf galaxies and globular clusters.  

High-resolution HST-images of the star forming regions in
interacting galaxies like the Antennae (Whitmore et al.\ 1999;
Zhang \& Fall 1999) or Stephan's Quintet (Gallagher et al.\ 2001)
resolve some of these regions into dozens to hundreds of young
massive star clusters. According to Whitmore et al.\ the
individual clusters have effective radii of about 4~pc and masses
$10^{4}$--$10^{6}$~M$_{\odot}$, with a mass-spectrum following a
steep power law $\Psi_{M} \propto M^{-2}$ (Zhang \& Fall 1999).
The striking point is that these young star clusters are
themselves clustered into groups of a few to several hundred star
clusters spanning projected regions of a few~100~pc, with a
higher cluster concentration at the centre.  Measurements of the
relative velocities between the star clusters within such objects
are now becoming available.  Preliminary results indicate
$\approx20$~km/s (B.~Whitmore, private communication), which is
consistent with virial masses $\approx10^7\,M_\odot$.  Age
determinations in the Antennae show that these star clusters are
extremely young (3--7~Myr).  While in the Antennae young massive
star clusters are preferably found in the central region of the
interacting pair, NGC~7319 in Stephan's Quintet has young star
clusters in the long tidal arm and the intra-group region north
of NGC~7319 (Gallagher et al.\ 2001).  


\section{Setup}
\label{sec:setup}

The orbital integration of the particles is performed with the 
particle-mesh code {\sc Superbox} (Fellhauer et al.\ 2000).  In
{\sc Superbox} densities are derived on Cartesian grids using the
nearest-grid-point (NGP) scheme.  From these density arrays the
potential is calculated via the fast Fourier-transformation.
Forces are obtained using higher-order differentiation based on
the NGP scheme but comparable in precision with standard CIC
(cloud-in-cell) algorithms.  The particles are integrated using a
fixed time-step Leap-Frog algorithm.  For an improved resolution
at the regions of interest, a hierarchical grid-architecture with
two levels of high resolution sub-grids are used for each star
cluster in the super-cluster.  These sub-grids track the density
maxima of the individual star clusters, and are adjusted at the
beginning of the computation to meet individual requirements. 

The young massive star clusters are modeled as Plummer-spheres
with Plummer-radii of 4~pc and cutoff-radii of 20~pc.  They have
masses of $10^{6}$~M$_{\odot}$ and $3.2 \cdot 10^{5}$~M$_{\odot}$
to mimic a mass spectrum.  32 star clusters (29 light and 3
heavy) are placed together in a sphere with places and relative
velocities according to a Plummer-distribution with a
Plummer-radius of 50~pc and a cutoff-radius of 250~pc.  This
configuration is than placed on an highly eccentric orbit around
a parent galaxy which is modeled as a rigid spherical potential.
The model starts at an apogalacticon position at 20~kpc and
passes perigalacticon at 2~kpc distance.  The total integration
time was 10~Gyr.

\section{The merger object}
\label{sec:object}

Within 500~Myr all 32 clusters have merged, building up a dense,
compact merger object.  The total mass of the merger object is
$\approx 10^{7}$~M$_{\odot}$ after formation.  But the object
looses mass with every perigalacticon passage.  Time-averaged the
best fit is a linear decrease of the mass.  After 10~Gyr the
object has lost about half of its mass but it still forms a
stable and bound object.  The surface density after 10~Gyr can be
well-fitted with a King profile with a tidal radius of about
400~pc and a core radius of about 11~pc.  There is a deviation
from a simple King profile in the range between the tidal radius
at perigalacticon ($\approx 150$~pc) and the actual tidal radius
but this is due to the fact that a simple King profile does not
account for a varying tidal field.  Fitting the different parts
of the profile shows that an exponential with a central surface
density of about 4500~M$_{\odot}/$pc$^{2}$ and an exponential
scale length of $12.5$~pc fits the very inner part best.  Out to
the tidal radius of the last perigalacticon a profile following a
$r^{-2}$ power law is the best choice of fitting.  The outermost
part is best fitted by a steep power-law proportional to
$r^{-4.5}$ out to the actual tidal radius.  Beyond that the
surface density turns over to a flat distribution of underlying
extra-tidal stars.  This can be understood by this
material gets unbound near perigalacticon.  But the stars do not
leave this region immediately and some are recaptured, when the
object is near apogalacticon.  The central space density of this
object is about 250~M$_{\odot}/$pc$^{3}$.  We also analyzed the
line-of-sight velocity dispersion distribution.  The central
velocity dispersion is about 12~km/s and follows an exponential
profile but with an exponential scale length of almost the size
of the system.  Measuring the line-of-sight velocity dispersion
reveals unbound stars even within the merger
object. This leads to an artificial rise of the velocity
dispersion in the outer parts (outside the core).  Even if stars
with a clear deviation from the velocity distribution of the
bound stars are removed, a rising velocity dispersion near the
tidal radius remains.

\section{Conclusions \& Outlook}
\label{sec:conclus}

Our models show that star cluster aggregates like the ones found
in the Antennae or Stephan's Quintet are very likely to merge
thereby building up merger objects which have sizes spanning from
massive globular clusters (like Omega Centaurus) to ultra-compact
dwarf galaxies (as found in the Fornax Cluster) up to very small
dwarf ellipticals.  This shows that interactions between gas-rich
galaxies followed by interaction-triggered star-formation
are possible origins for second-generation dwarf
galaxies.  The main difference between dwarf galaxies generated
by such events and primordial dwarfs is that they are not
dark matter dominated.  In the young universe the
interaction rate between galaxies must have been higher,
with ubignitous formation of 
second-generation dwarf galaxies.  Our simulations show
that even the smaller ones are able to survive in a strong and
varying tidal field for more than a Hubble time.  

From the observational point of view one can discriminate
between the formation scenario as proposed in this project and
the primordial one by measuring the dark matter content of dwarf
objects.  

\acknowledgements
MF acknowledges financial support through DFG-grant FE564/1-1.

\end{article}

\begin{thebibliography}{}
\bibitem{} Fellhauer M., Kroupa P., Baumgardt H., Bien R.,   
  Boily C.M., Spurzem R., Wassmer N., 2000,  NewA, {\bf 5},
  305
\bibitem{} Fellhauer M., Baumgardt H., Kroupa P., Spurzem R.,
  2001, to appear in Cel.\ Mech.\ \& Dyn.\ Astron.,
  astro-ph/0103052 
\bibitem{} Fellhauer M., Kroupa P., 2001, to appear in MNRAS,
  astro-ph/0110621 
\bibitem{} Gallagher S.C., Charlton J.C., Hunsberger S.D.,
  Zaritsky D., Whitmore B.C., 2001, AJ, in press
  (astro-ph/0104005)
\bibitem{} Hilker M., Infante L., Kissler-Patig M., Richtler T.,
  1999, A\&AS, {\bf 134}, 75
\bibitem{} Kroupa P., 1997, NewA, {\bf 2}, 139 
\bibitem{} Mieske S., Hilker M., 2001, AG Abstract Series, {\bf
    18}, 215 
\bibitem{} Phillipps S., Drinkwater M.J., Gregg M.D., Jones
  J.B., 2001, to appear in ApJ, astro-ph/0106377
\bibitem{} Whitmore B.C., Zhang Q., Leitherer C., Fall S.M.,
  1999, AJ, {\bf 118}, 1551 
\bibitem{} Zhang Q., Fall S.M., 1999,  ApJL, {\bf 527}, 81L
\end{thebibliography}
\end{document}